\documentclass[
 reprint,
nofootinbib,
 amsmath,amssymb,
 aps,
 prd,
]{revtex4-2}

\usepackage{dcolumn}
\usepackage{indentfirst}
\usepackage{amsmath}
\usepackage{bm}
\usepackage{slashed}
\usepackage{graphicx}
\usepackage{hyperref}
\usepackage{multirow}
\usepackage{adjustbox}

\begin{document}

\preprint{APS/123-QED}

\title{Interacting scalar field dark matter and stepped dark radiation in an extended Wess-Zumino dark radiation model}

\author{Gang Liu}
 \email{liugang@hlju.edu.cn}

\affiliation{
 College of Physical Science and Technology\\
 Heilongjiang University\\
 Harbin 150080, People's Republic of China
}

\date{\today}

\begin{abstract}
    We extend the Wess-Zumino dark radiation (WZDR) model by replacing cold dark matter with scalar field dark 
    matter and introducing a pure momentum coupling between the scalar field and stepped dark radiation. This 
    coupling avoids excessive early-universe damping, thereby preserving the predictive power of the WZDR 
    framework while offering a potential mechanism for exploring the Hubble and $S_8$ tensions.
    Using a combination of cosmic microwave background, baryon acoustic oscillations, Type Ia supernovae, the 
    SH0ES $H_0$ measurement, Dark Energy Survey Year 3 (DES) $S_8$ data, and Atacama Cosmology Telescope data, we 
    constrain the model. When only one of the SH0ES and DES priors is included, the new model outperforms WZDR 
    in $\chi^2$ and $Q_{\rm DMAP}$; with both priors, it is disfavored. Using the latest cosmological data, we 
    find that at 68\% C.L. the coupled model gives $H_0 = 71.08^{+0.73}_{-0.66}$ km/s/Mpc and $S_8 = 0.8162\pm0.0073$, 
    close to the WZDR values of $71.20^{+0.73}_{-0.61}$ and $0.8165^{+0.0078}_{-0.0066}$. Both models raise 
    $H_0$ effectively, but the new model only marginally reduces $S_8$, which remains above the $\Lambda$CDM 
    prediction. The coupling is weak, with $\log_{10}(\xi) < -3.36$ (95\% C.L.), and performs worse than WZDR 
    in both $\chi^2$ and AIC. Hence, although it does not fully resolve the tensions, this work provides a 
    useful reference for studies of interacting dark sectors. 
\end{abstract}

\maketitle


\section{Introduction}

In recent years, with the increasing quantity and improved precision of cosmological observations, several 
cosmological tensions have become more pronounced \cite{ABDALLA202249}, with the most prominent being the 
Hubble tension and the $S_8$ tension.

The Hubble constant, $H_0$, is a fundamental cosmological parameter that characterizes the current rate of 
expansion of the universe. According to Cosmic Microwave Background (CMB) observations, the \textit{Planck} 
Collaboration (2018) inferred a value for the Hubble constant of $67.37\pm0.54$ km/s/Mpc within the context 
of the $\Lambda$CDM model \cite{planck2020}. In contrast, the Supernova $H_0$ for the Equation 
of State (SH0ES) team obtained $H_0=73.04\pm1.04$ km/s/Mpc from Cepheid-calibrated Type Ia supernovae (SNIa) 
measurements \cite{Riess_2022}. This discrepancy of approximately 4.8$\sigma$ between the two results 
constitutes the Hubble tension \cite{Verde_2019}.
The Hubble tension has now become one of the most significant challenges in modern cosmology, for comprehensive 
overviews of the problem and its proposed solutions, see Refs.~\cite{2103.01183,2211.04492}.  

The $S_8$ tension refers to the discrepancy between the value of $S_8$ derived from CMB measurements and 
that obtained from weak lensing and galaxy count observations \cite{planck8,kids1}. The \textit{Planck} 2018 
best-fit $\Lambda$CDM model yields a value of $0.834\pm0.016$ \cite{planck2020}. However, 
large-scale structure observations report smaller values, such as $0.759^{+0.024}_{-0.021}$ from KiDS-1000 
\cite{KiDS}, and $0.776\pm0.017$ from the Dark Energy Survey Year 3 \cite{PhysRevD.105.023520}.
The systematic status of the $S_8$ tension, including the persistent discrepancies among different large-scale 
structure probes, has been comprehensively examined in the literature \cite{2008.11285,2602.12238}.

Recent analyses of the KiDS-Legacy final data release have claimed $S_8$ consistent with \textit{Planck}, 
suggesting the tension may disappear within $\Lambda$CDM \cite{2025arXiv250319442S}. However, other surveys 
(DES Y3, KiDS-1000) continue to find lower $S_8$, while ACT and \textit{Planck} favor higher $S_8$. The split 
in observational constraints has deepened, not resolved. Crucially, this consistency analysis is confined to 
$\Lambda$CDM. Therefore, investigating the $S_8$ tension within cosmological models extending beyond $\Lambda$CDM 
remains of considerable importance.

A relatively promising solution to the Hubble tension involves reducing the acoustic horizon prior to 
recombination \cite{Aylor_2019}, which facilitates an increase in the value of $H_0$ without conflicting with 
the CMB observations. This approach is primarily represented by two models: the early dark energy model 
\cite{PhysRevLett.122.221301, AGRAWAL2023101347, PhysRevD.100.063542, POULIN2023101348, PhysRevD.103.L041303} 
and the additional radiation model \cite{Bernal_2016, Choudhury_2021, PhysRevD.104.063523, PhysRevD.101.123505}. 
The early dark energy model introduces a new fluid whose energy density evolves over time, and, through 
fine-tuning, it affects the system only prior to recombination, after which it rapidly vanishes.

The additional radiation model introduces a dark radiation component that avoids the need for fine-tuning, as 
the radiation energy density naturally redshifts following the matter-radiation equality. These models 
incorporate both free-streaming and self-interacting dark radiation \cite{Brust_2017, Blinov_2020, 
Escudero_2021}.

However, these models continue to face several challenges. For instance, while early dark energy models 
alleviate the Hubble tension, they often exacerbate the $S_8$ tension \cite{PhysRevD.102.043507, D'Amico_2021, 
PhysRevD.109.103531}. The inclusion of additional radiation can enhance Silk damping \cite{PhysRevD.87.083008}, 
which suppresses the high-multipole power spectrum of the CMB, thereby conflicting with high-$\ell$ polarization 
data \cite{Bernal_2016, Choudhury_2021, PhysRevD.104.063523, Blinov_2020}.

In parallel with early-universe modifications, the coupled dark sector scenario with non-gravitational energy 
exchange between dark energy and dark matter has drawn growing interest. Motivated by its long theoretical 
foundation \cite{astro-ph/9908023}, recent DESI/\textit{Planck} analyses indicate a preference for such interactions 
at more than 95\% confidence level \cite{2404.15232}, with further studies reporting 3-5$\sigma$ significance 
\cite{2407.14934,2601.07361}.

An alternative 
viewpoint suggests that if a physical theory can fix extra parameters to specific nonstandard values, the 
resulting model may evade the Bayesian prior penalty typically associated with parameter extensions 
\cite{1907.07569}.

It has recently been argued that early-time new physics alone may be insufficient 
to fully resolve the Hubble tension, suggesting that a combination of early- and late-time modifications might 
be necessary \cite{2308.16628,bc81-3fj5}. 

Motivated by these results, we consider the stepped dark radiation model as an 
early-universe new physics scenario and introduce its interactions with dark matter, aiming to examine whether 
such a combination can mitigate the cosmological tensions.

The stepped dark radiation model, specifically the Wess-Zumino dark radiation (WZDR) model \cite{PhysRevD.105.123516}, 
effectively addresses many of the aforementioned issues. This model is grounded in the well-motivated and 
straightforward Wess-Zumino supersymmetry theory. It not only alleviates the Hubble tension but also contributes 
to the mitigation of the $S_8$ tension \cite{PhysRevD.108.023520}.

Building on the stepped dark radiation model, extensive research has been conducted \cite{PhysRevD.107.063536, 
Allali_2024}, with particular emphasis on introducing interactions between cold dark matter and stepped dark radiation. These 
studies investigate the potential of coupled models to simultaneously alleviate both the Hubble tension and the 
$S_8$ tension \cite{Zhou:2024igb, Schöneberg_2022, PhysRevD.108.123513, PhysRevD.111.043513}. 

In the present work, we focus on the pure momentum coupling between scalar field dark matter (SFDM) and stepped 
dark radiation. Compared to other extensions of the WZDR model, this coupling transfers no energy at the 
background level, operating only at the perturbation level: it alters the relative velocity between dark matter 
and dark radiation without modifying the background densities. This avoids excessive early-universe damping, 
while adjusting late-time structure formation to address tensions such as $S_8$.

SFDM, composed of light scalar fields with a mass of approximately $10^{-22}$ eV, is a promising candidate for 
dark matter \cite{Ferreira_2021, PhysRevD.106.123501}. On small scales, it can form condensates that suppress 
structure growth, while on larger scales, it behaves similarly to cold dark matter. As a result, it may play a 
role in alleviating the $S_8$ tension \cite{PhysRevD.108.083523, liu2023kinetically}.

\begin{figure}
  \centering
	\includegraphics[width=.9\columnwidth]{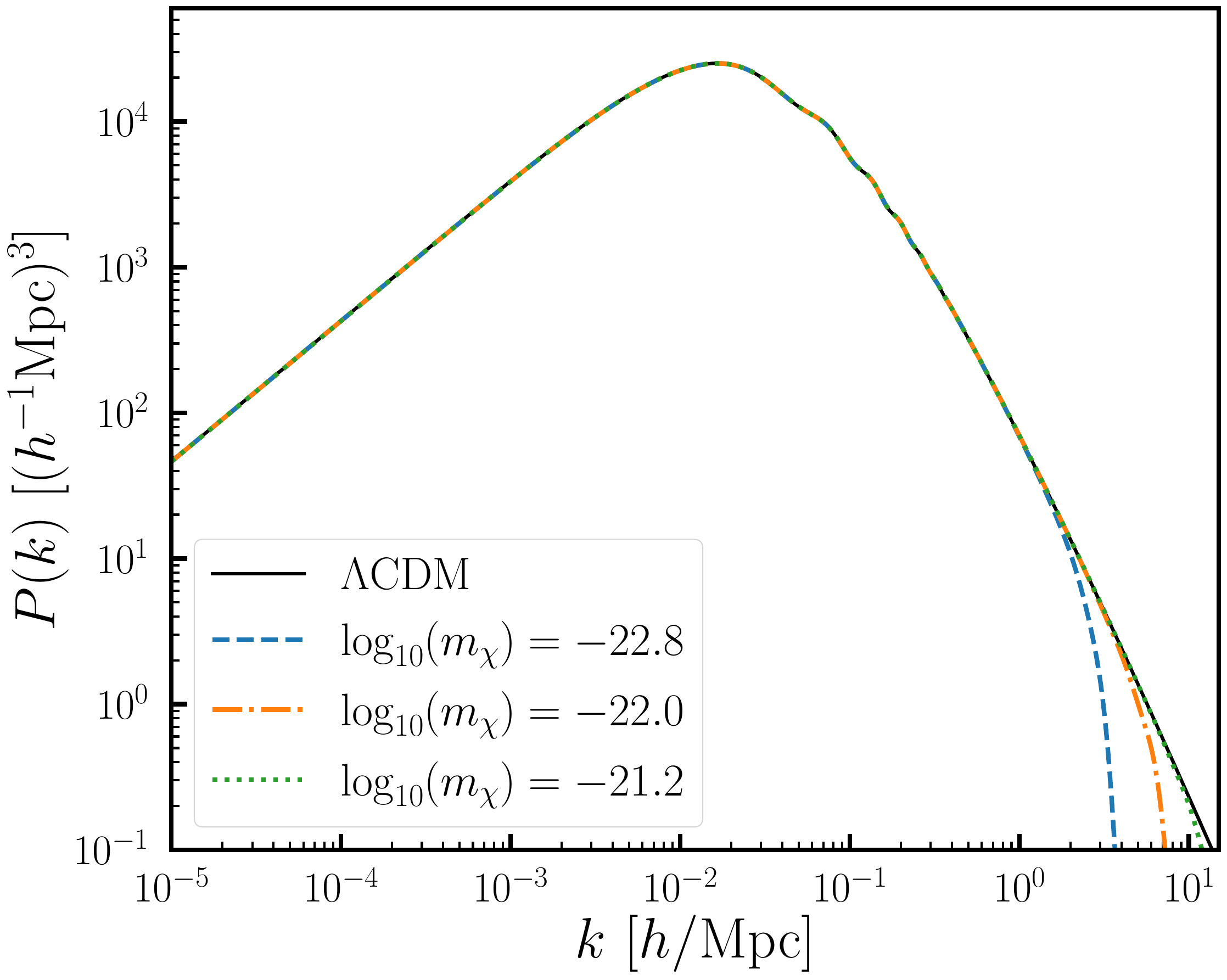}
  \caption{The impact of varying the SFDM mass on the linear matter power spectrum is investigated. On small 
  scales, the power spectrum for SFDM is suppressed, whereas on large scales, the results are in agreement 
  with those of the $\Lambda$CDM model.}
  \label{fig:1}
\end{figure}

Figure \ref{fig:1} illustrates the effect of varying the mass of SFDM on the linear matter power spectrum, with 
all other cosmological parameters set to values consistent with the $\Lambda$CDM model. It can be observed that, 
on small scales, the matter power spectrum for SFDM is suppressed relative to that of the $\Lambda$CDM model, 
whereas on large scales, the results converge with those of the $\Lambda$CDM model.

Previous studies on pure momentum interactions in the dark sector are primarily based on various coupled 
quintessence models, which describe the interaction between dark matter and dark energy \cite{PhysRevD.88.083505, 
PhysRevD.94.043518, PhysRevD.101.043531}. In the present work, we treat the stepped dark radiation as a perfect 
fluid and, building on the generalized fluid action from Ref.~\cite{PhysRevD.101.043531}, propose a novel WZDR+ 
model to describe the interaction between SFDM and the stepped fluid. 

Admittedly, this phenomenological fluid treatment requires further field-theoretic justification. However, for 
large-scale structure evolution, approximating dark radiation as a perfect fluid is a well-established approach 
(see, e.g., Refs.~\cite{Allali_2025,PhysRevD.108.123513}). We therefore adopt it as a practical first step to 
explore the observable consequences of pure momentum coupling. A more fundamental derivation from a 
field-theoretic framework is left for future work.

This paper is organized as follows. In Section~\ref{sec:model}, we introduce the novel coupling model and derive 
the background and perturbation evolution equations for the dark sectors in detail. Section~\ref{sec:nr} 
presents numerical analyses illustrating the effects of coupling on both the matter power spectrum and the CMB 
power spectrum. In Section~\ref{sec:mc}, we describe the datasets used for the Markov Chain Monte Carlo (MCMC) 
analysis and present the resulting parameter constraints. Finally, in Section~\ref{sec:con}, we discuss and 
summarize our findings.

\section{Coupling model}
\label{sec:model}
We define the following operator to describe the interaction between SFDM and the dark radiation fluid \cite{PhysRevD.101.043531}: 
\begin{equation}
    \mathcal{L}_\mathrm{int}=\xi(u^{\mu}\partial_{\mu}\chi)^2,
\end{equation}
where $\xi$ is a dimensionless coupling constant that characterizes the strength of the interaction, 
$u^{\mu}$ represents the four-velocity of the dark fluid, and $\chi$ denotes the scalar field. 
In this framework, we focus exclusively on the simple quadratic form of the scalar field potential, 
\begin{equation}
    V(\chi)=\frac{1}{2}m^2\chi^2,
\end{equation}
as more complex potential forms can generally be expressed in terms of this quadratic form \cite{Ureña-López_2016}. 

By applying the calculus of variations, we can derive the total energy-momentum tensor for SFDM and dark radiation 
as follows:
\begin{equation}
    \begin{aligned}
        T^{\mu}_{(\chi)\nu}&+T^{\mu}_{(d)\nu}
        =-2\xi(u^{\alpha}\partial_{\alpha}\chi)(u^{\mu}\partial_{\nu}\chi)\\
        &+\partial^{\mu}\chi\partial_{\nu}\chi+(\rho_d+p_d)u^{\mu}u_{\nu}+p_d\delta^{\mu}{}_{\nu}\\
        &-\delta^{\mu}{}_{\nu}\left[\frac{1}{2}\partial^{\alpha}\chi\partial_{\alpha}\chi+V(\chi)
        -\xi(u^{\alpha}\partial_{\alpha}\chi)^2\right],
    \end{aligned}
    \label{eqT}
\end{equation}
where the subscript ``$d$'' denotes the dark radiation fluid, with $\rho_d$ and $p_d$ representing the energy 
density and pressure of the dark fluid, respectively. 

We assume that these two dark components do not interact with any other components, except through gravity. 
By invoking the covariant conservation of the total energy-momentum tensor, we derive the following relationship, 
\begin{equation}
    \nabla_{\mu}T^{\mu\nu}_{(\chi)}=-\nabla_{\mu}T^{\mu\nu}_{(d)}.
    \label{eqC}
\end{equation}

\subsection{Background equations}
The background evolution equations for SFDM and dark radiation can be derived using the calculus of variations, 
in conjunction with Eq.~(\ref{eqC}),
\begin{subequations}
    \begin{align}
        \ddot{\chi}+3H\chi+\frac{m^2\chi}{1+2\xi}&=0, \label{kg}\\
        \dot{\rho_d}+3H(\rho_d+p_d)&=0,
    \end{align} 
\end{subequations}
where the dot denotes the derivative with respect to cosmic time, and $H$ is the Hubble parameter. 

It can be observed that the evolution equation for the dark radiation fluid remains unchanged. 
Therefore, at the background level, we adopt the same description method as in the original WZDR model to 
treat the dark fluid \cite{PhysRevD.105.123516}.

The energy density and pressure of SFDM, accounting for interactions between dark components, can be expressed 
in the following form, 
\begin{subequations}
    \begin{align}
        \rho_{\chi}&=\frac{1}{2}\dot{\chi}^2(1+2\xi)+\frac{1}{2}m^2\chi^2,\\
        p_{\chi}&=\frac{1}{2}\dot{\chi}^2(1+2\xi)-\frac{1}{2}m^2\chi^2.
    \end{align} 
    \label{rhod}
\end{subequations}
Based on Eqs.~(\ref{kg}) and (\ref{rhod}), one can deduce that the model is physically viable for $\xi>-\frac{1}{2}$. 
As $\xi \to -\frac{1}{2}$, a strong coupling problem arises. For $\xi < -\frac{1}{2}$, the negative kinetic 
term leads to the appearance of ghost in the theory.

We introduce new variables to derive the equation of motion for the scalar field \cite{Ureña-López_2016,
PhysRevD.96.061301,PhysRevD.57.4686}, 
\begin{subequations}
    \begin{align}
        \sqrt{\Omega_{\chi}}\sin{\frac{\alpha}{2}}&=\frac{\dot{\chi}\sqrt{1+2\xi}}{\sqrt{6}M_{pl}H},\\
        \sqrt{\Omega_{\chi}}\cos{\frac{\alpha}{2}}&=-\frac{m\chi}{\sqrt{6}M_{pl}H},\\
        y_1&=\frac{2m}{H},
    \end{align} 
\end{subequations}
where $\Omega_{\chi}$ represents the energy density fraction of dark matter 
and $M_{pl}$ denotes the reduced Planck mass. Using the Klein-Gordon equation for SFDM in 
Eq.~(\ref{kg}), we obtain the evolution equation for the new variables as follows, 
\begin{subequations}
    \begin{align}
        \frac{\dot{\Omega}_{\chi}}{\Omega_{\chi}}&=3H(w_t+\cos\alpha),\\
        \dot{\alpha}&=-3H\sin\alpha+\frac{Hy_1}{\sqrt{1+2\xi}},\\
        \dot{y_1}&=\frac{3}{2}H(1+w_t)y_1,
    \end{align} 
\end{subequations}
where $w_t$ represents the total equation of state, defined as the ratio of total pressure to total energy 
density of the cosmic components.

\subsection{Perturbation equations}
We examine the perturbation equations for SFDM and dark radiation in the synchronous gauge, where the line 
element is given by:
\begin{equation}
    \mathrm{d}s^2=-\mathrm{d}t^2+a^2(t)(\delta_{ij}+h_{ij})\mathrm{d}x^i\mathrm{d}x^j.
\end{equation}
The perturbed Klein-Gordon equation for SFDM in Fourier space is expressed as: 
\begin{small}
\begin{equation}
    \ddot{\delta\chi}+3H\dot{\delta\chi}+\frac{1}{2}\dot{h}\dot{\chi}
    +\frac{(k^2+a^2m^2)\delta\chi}{a^2(1+2\xi)}+\frac{2\xi\dot{\chi}\theta_d}{a(1+2\xi)}=0,
    \label{pkg}
\end{equation}
\end{small}
where $h$ denotes the trace of the scalar metric perturbations. By combining this with the covariant 
conservation equation of the total energy-momentum tensor, we can derive the evolution equations for the energy 
density perturbation and velocity of the dark fluid, 
\begin{equation}
    \dot{\delta_d}+3H(c^2_s-w_d)\delta_d=-(\frac{1}{2}\dot{h}+\frac{\theta_d}{a})(1+w_d),
\end{equation}
\begin{equation}
    \begin{aligned}
        \dot{\theta_d}+H\theta_d(1-3c^2_s)&=\frac{k^2c^2_s\delta_d}{a(1+w_d)}\\
        &+\frac{2H\xi\dot{\chi}}{\rho_d(1+w_d)}(\frac{k^2}{a}\delta\chi-\dot{\chi}\theta_d).
    \end{aligned}
    \label{thetad}
\end{equation}
Here, $\delta_d$ and $\theta_d$ represent the energy density contrast and velocity divergence of the dark 
radiation fluid, respectively, $w_d$ is the equation of state of the dark fluid, and $c_s$ 
denotes the sound speed of the fluid (see Ref.~\cite{PhysRevD.105.123516} for their definitions).

By examining the density perturbation equation and the velocity divergence equation of dark radiation, 
it is evident that the former retains its original form in the absence of coupling, with modifications occurring 
solely in the velocity evolution equation of the dark fluid.

The energy density perturbation, pressure perturbation, and velocity divergence of SFDM can be expressed in the 
following form \cite{PhysRevD.58.023503,Hu_1998}: 
\begin{subequations}
    \begin{align}
        \delta\rho_{\chi}&=\dot{\chi}\dot{\delta\chi}(1+2\xi)+m^2\chi\delta\chi,\\
        \delta p_{\chi}&=\dot{\chi}\dot{\delta\chi}(1+2\xi)-m^2\chi\delta\chi,\\
        (\rho_{\chi}+p_{\chi})\theta_{\chi}&=\frac{k^2}{a}\dot{\chi}\delta\chi(1+2\xi).
    \end{align} 
\end{subequations}
Equation~(\ref{thetad}) can therefore be rewritten in the following form: 
\begin{small}
\begin{equation}
    \begin{aligned}
    \dot{\theta_d}+H\theta_d(1-3c^2_s)&=\frac{k^2c^2_s\delta_d}{a(1+w_d)}\\
    &+\frac{2H\xi}{1+2\xi}\cdot\frac{\rho_\chi(1-\cos\alpha)}{\rho_d(1+w_d)}(\theta_\chi-\theta_d),
\end{aligned}
\end{equation}
\end{small}
from which the momentum coupling between SFDM and the dark fluid is more clearly revealed.

We proceed by introducing new variables to derive the perturbation equations for the scalar field, 
\begin{subequations}
    \begin{align}
        \sqrt{\Omega_\chi}\left(\delta_0\sin\frac{\alpha}{2}+\delta_1\cos\frac{\alpha}{2}\right)
        &=\sqrt{\frac{2}{3}}\frac{\sqrt{1+2\xi}}{M_{pl}H}\dot{\delta\chi},\\
        \sqrt{\Omega_\chi}\left(\delta_1\sin\frac{\alpha}{2}-\delta_0\cos\frac{\alpha}{2}\right)
        &=\sqrt{\frac{2}{3}}\frac{m\delta\chi}{M_{pl}H}.
    \end{align} 
\end{subequations}
The evolution equation for the new variables is expressed as: 
\begin{equation}
    \begin{aligned}
        \dot{\delta_{0}}=\delta_0H\omega\sin\alpha&-\delta_1[3H\sin\alpha+H\omega(1-\cos\alpha)]\\
        &-\left[\frac{2\xi\theta_d}{(1+2\xi)a}+\frac{\dot{h}}{2}\right](1-\cos\alpha),\\
    \end{aligned}
\end{equation}
\begin{equation}
    \begin{aligned}
        \dot{\delta_{1}}=\delta_0H\omega(1+\cos\alpha)&-\delta_1(3H\cos\alpha+H\omega\sin\alpha)\\
        &-\left[\frac{2\xi\theta_d}{(1+2\xi)a}+\frac{\dot{h}}{2}\right]\sin\alpha,
    \end{aligned}
\end{equation}
where 
\begin{equation}
    \omega=\frac{k^2}{a^2H^2y_1}\cdot\frac{1}{\sqrt{1+2\xi}}.
\end{equation}

\subsection{Initial conditions}
In the very early universe, Hubble friction effectively froze the scalar field at its initial value, resulting 
in a slow-roll process. This allows the initial value of $\dot{\chi}$ to be set to 0. 
As a consequence, the evolution equations for the dark radiation fluid reduce to their uncoupled forms. Thus, 
for the dark fluid, we retain the initial condition settings from the original WZDR model. For the SFDM, we adopt 
and modify the attractor solution initial condition \cite{Ureña-López_2016, PhysRevD.96.061301}: 
\begin{subequations}
    \begin{align}
        \alpha_{i}&=\frac{2}{5}\frac{m}{H_0\sqrt{\Omega_r a_{i}^{-4}}}\cdot\frac{1}{\sqrt{1+2\xi}},\\
        y_{1i}&=5\sqrt{1+2\xi}\alpha_{i},
    \end{align} 
\end{subequations}
where $\Omega_r$ denotes the present energy density fraction of the radiation component, and $a_i$ represents 
the initial value of the scale factor. The initial value of $\Omega_\chi$ can be determined from its current 
value using the shooting method within the Boltzmann code \texttt{CLASS}\footnote{\url{https://github.com/lesgourg/class_public}} \cite{1104.2932,Blas_2011}. 

\section{Numerical results}
\label{sec:nr}
Based on the description in the previous section, we modified the publicly available Boltzmann code \texttt{CLASS} 
to obtain numerical results. Cold dark matter was replaced with SFDM, and to maintain consistency with the 
original code's conventions, we set the energy density fraction of cold dark matter, $\Omega_{\mathrm{cdm},0}$, 
to $10^{-6}$ \cite{Ureña-López_2016}.

We use the cosmological parameter values from \cite{PhysRevD.105.123516} to present the numerical 
results. Specifically, for the WZDR model, we adopt the following values for the parameters: 
\begin{align}
    &H_0=69.3, \quad \ln(10^{10}A_\mathrm{s})=3.050,\\
    \notag
    &\omega_\mathrm{b}=0.02253, \quad \tau_\mathrm{reio}=0.0577,\\
    \notag
    &\omega_\mathrm{dm}=0.12390, \quad n_\mathrm{s}=0.9721,\\
    \notag
    &N_\mathrm{IR}=0.28, \quad \log_{10}(z_\mathrm{t})=4.29.
\end{align}

For the WZDR+ model, we fix the SFDM mass at $10^{-22}$ eV and vary only the coupling parameter $\xi$, while 
keeping the other parameters consistent with those in the WZDR model.

\begin{figure}
    \centering
	\includegraphics[width=.9\columnwidth]{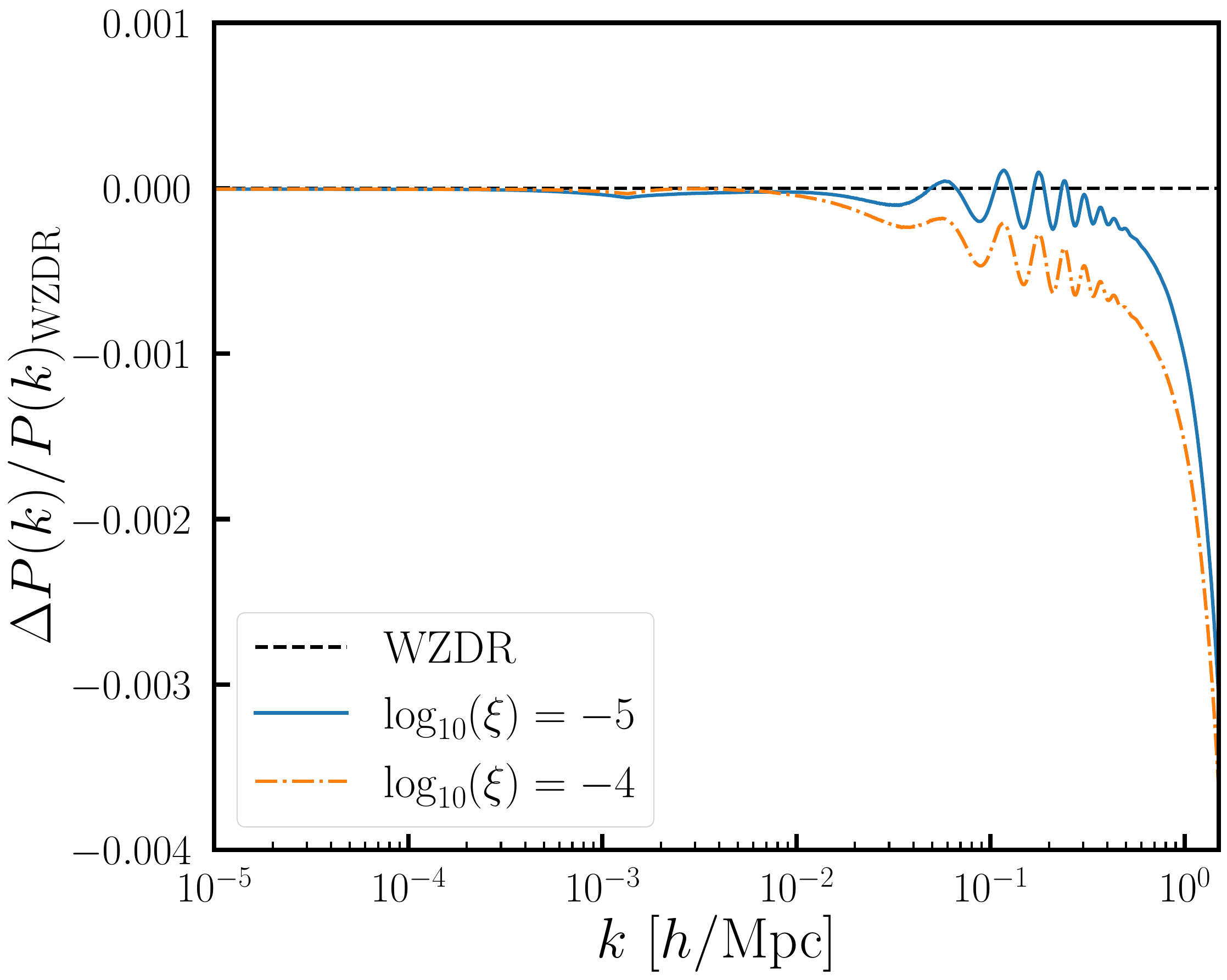}
    \caption{The linear matter power spectrum of the WZDR+ model exhibits a reduction on small scales relative 
    to the original WZDR model. This suppression results from the combined effects of the intrinsic property of 
    SFDM and the momentum exchange between dark matter and dark radiation.}
    \label{fig:2}
\end{figure}

In Fig.~\ref{fig:2}, we present the variation in the linear matter power spectrum of the WZDR+ model relative to 
the original WZDR model. 
The black dashed line represents the results of the WZDR model, while the blue solid line and the orange 
dashed-dotted line correspond to the results for coupling constant $\xi=10^{-5}$ and $\xi=10^{-4}$ in the 
new model, respectively.

We now specify the parameter $\xi$. The background equations in Section~\ref{sec:model} require 
$\xi>-\frac{1}{2}$ for stability. In fact, the coupling structure in Eqs.~(\ref{pkg}) and (\ref{thetad}) 
further restricts to $\xi>0$, since only in this regime does the SFDM density contrast acquire a friction 
term and the dark radiation velocity gain a source term, which together suppress structure growth. 
We adopt $\xi=10^{-5}$ and $\xi=10^{-4}$ for numerical demonstrations. 

The power spectrum of the WZDR+ model on small scales is found to be smaller than that of the WZDR model, 
exhibiting an oscillatory behavior. This result is attributed to the combined effects of the properties of 
SFDM and the interaction between dark matter and dark radiation.

First, the matter power spectrum of SFDM is suppressed below the ``Jeans scale'', accompanied by regular damped 
oscillations, which constitute a distinctive feature of the SFDM model \cite{Ureña-López_2016}. Second, the 
pure momentum interaction between dark matter and dark radiation indirectly affects the evolution of density 
perturbations by modifying the velocity evolution of dark radiation. This interaction further suppresses the 
matter power spectrum on small scales, a mechanism that may help alleviate the $S_8$ tension. Moreover, the 
oscillatory effects induced by the relatively weak interaction couple with the intrinsic oscillations from SFDM, 
resulting in an overlapping oscillatory pattern.

\begin{figure}
    \centering
	\includegraphics[width=.9\columnwidth]{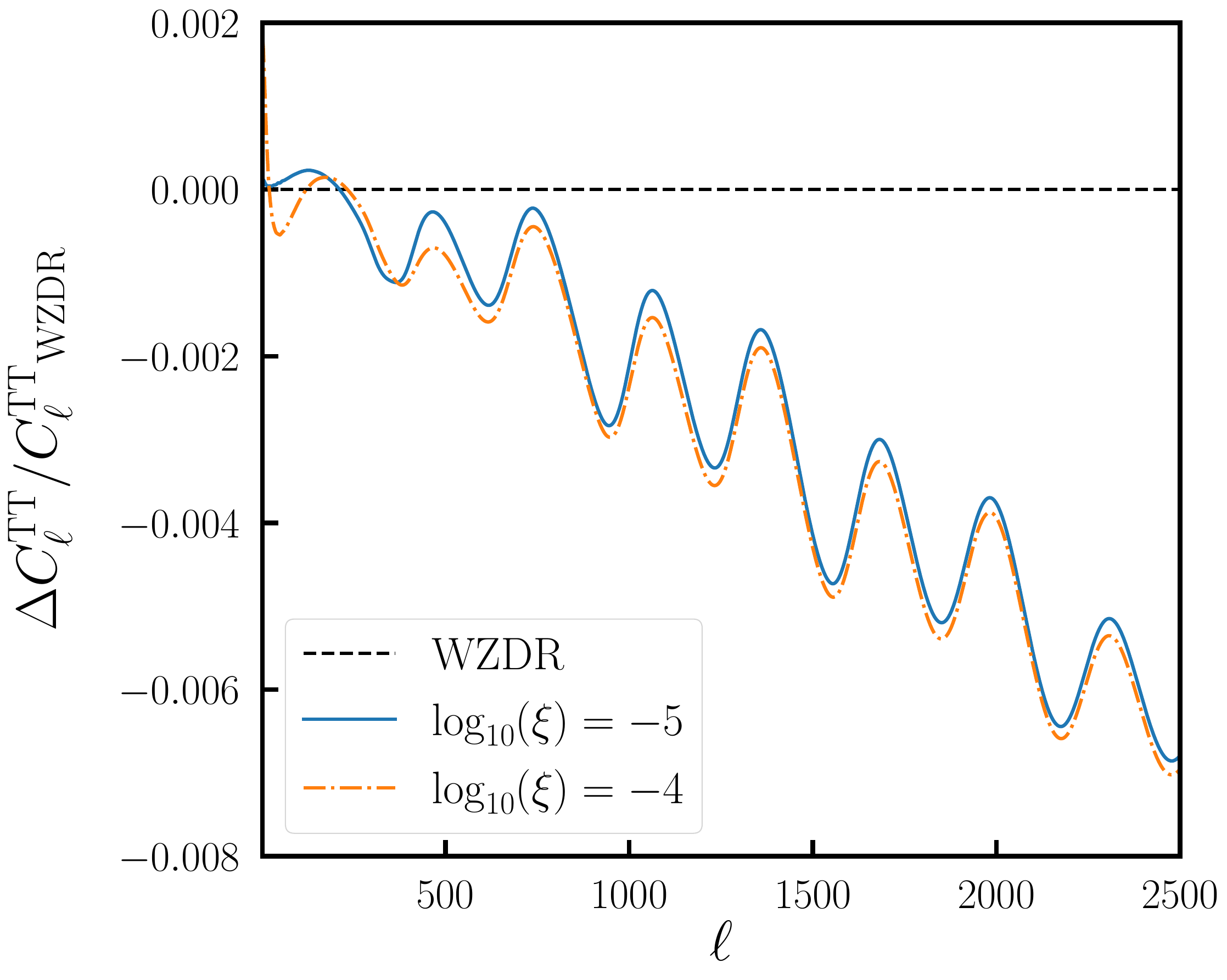}
    \caption{The differences in the CMB temperature power spectrum between the new model with varying coupling 
    constant and the WZDR model can be attributed to two main factors: the decay of gravitational potentials 
    during acoustic oscillations and the attenuation of gravitational lensing. Together, these effects 
    contribute to the overall reduction in the amplitude of the power spectrum.}
    \label{fig:3}
\end{figure}

Figure~\ref{fig:3} shows the differences in the CMB temperature power spectrum between the WZDR+ model with 
varying values of the coupling constant and the baseline WZDR model. At high multipoles, the amplitude of the 
power spectrum in the WZDR+ model is systematically suppressed. This suppression arises because momentum 
exchange between dark matter and dark radiation induces the decay of gravitational potentials during acoustic 
oscillations. As photons climb out of these decaying potential wells, they experience additional redshift, 
leading to enhanced damping in the CMB temperature anisotropies.

Furthermore, the dark matter-dark radiation coupling suppresses the matter power spectrum on small scales, 
reducing the number of structures capable of gravitational lensing. As a result, the gravitational lensing of 
CMB photons by large-scale structures is weakened, leading to CMB power spectrum peaks and troughs that are 
less smoothed by lensing compared to the predictions of the original model.

\section{MCMC analyses}
\label{sec:mc}
We performed MCMC analysis using \texttt{Cobaya}\footnote{\url{https://github.com/CobayaSampler/cobaya}} \cite{Torrado_2021} 
to obtain the posterior distribution of 
the model parameters, assessing the convergence of the results using the Gelman-Rubin criterion \cite{Gelman1992InferenceFI}, 
with $R-1<0.01$. Subsequently, we employed \texttt{GetDist}\footnote{\url{https://github.com/cmbant/getdist}} \cite{lewis2019getdist} 
to analyze the MCMC chains.

\begin{table}
    \caption{Prior ranges for the sampled cosmological parameters, including the six $\Lambda$CDM parameters, 
    the two WZDR parameters, and the coupling parameter.}
    \label{tab:1}
    \renewcommand{\arraystretch}{1.2}
    \begin{adjustbox}{max width=.9\linewidth}
    \begin{tabular} { l c c}
       \hline
       Parameter  & & Prior\\
       \hline
       $\ln(10^{10}A_\mathrm{s})$ & & [$1.61, 3.91$]\\
       $n_\mathrm{s}$ & & [$0.8, 1.2$]\\
       $\tau_\mathrm{reio}$ & & [$0.01, 1.8$]\\
       $H_0$ & & [$65, 80$]\\
       $\Omega_\mathrm{b}h^2$ & & [$0.005, 0.1$]\\
       $\Omega_\mathrm{dm}h^2$ & & [$0.001, 0.99$]\\
       $N_\mathrm{IR}$ & & [$0.0, 0.8$]\\
       $\log_{10}(z_\mathrm{t})$ & & [$4.0, 4.6$]\\
       $\log_{10}(\xi)$ & & [$-7.8, -1.0$]\\
       \hline         
    \end{tabular}
  \end{adjustbox}
  \end{table}

For parameter sampling in the WZDR+ model, we use the six standard $\Lambda$CDM model parameters \{$H_0$, 
$\Omega_\mathrm{b}h^2$, $\Omega_\mathrm{dm}h^2$, $\ln(10^{10}A_\mathrm{s})$, $n_\mathrm{s}$, $\tau_\mathrm{reio}$\}, 
along with the additional WZDR parameters \{$N_\mathrm{IR}$, $\log_{10}(z_\mathrm{t})$\} and the coupling 
constant $\xi$. 

The prior ranges for all sampled parameters are listed in Table~\ref{tab:1}. The six standard $\Lambda$CDM 
parameters adopt the default priors in \texttt{Cobaya}, while the two WZDR parameters follow the ranges given in Refs.~
\cite{PhysRevD.105.123516,PhysRevD.108.023520}. 
For the coupling parameter $\xi$, we conducted a preliminary MCMC analysis with \texttt{MontePython}\footnote{\url
{https://github.com/brinckmann/montepython_public}} \cite{brinck,Audren_2013} using standard cosmological 
data, adopting a prior mean of zero without imposing hard bounds. The posterior is of order $10^{-5}$. 
Moreover, values larger than about $0.1$ would trigger strong coupling and cause numerical issues in our code. 
We therefore impose $\log_{10}(\xi) \in [-7.8, -1.0]$.

\begin{figure}
    \centering
	\includegraphics[width=.9\columnwidth]{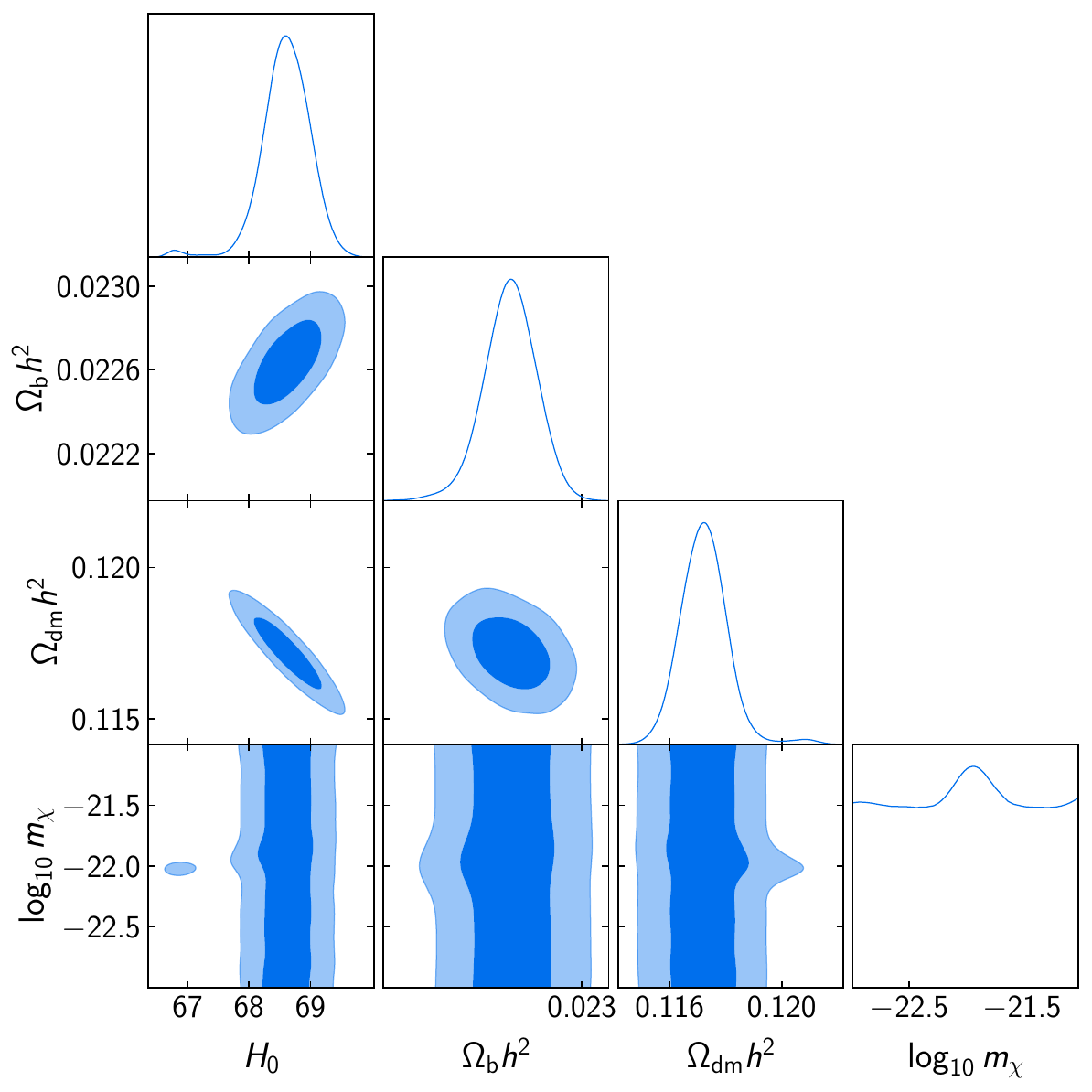}
    \caption{The MCMC results for the SFDM model (i.e., $\Lambda$CDM with cold dark matter replaced by SFDM), 
    obtained with the combined CMB, BAO, SNIa, SH0ES, and DES datasets (hereafter $\mathcal{DHS}$), show that 
    the current data cannot constrain the SFDM mass.}
    \label{fig:4}
\end{figure}

Based on our previous research, current commonly used cosmological datasets are unable to constrain 
the mass of SFDM \cite{PhysRevD.108.123546}. 
Using \texttt{Cobaya} with the full combination of CMB, BAO, SNIa, SH0ES, and DES data (hereafter denoted as 
$\mathcal{DHS}$), we present in Fig.~\ref{fig:4} 
some MCMC results for the SFDM model (i.e., $\Lambda$CDM with cold dark matter replaced by SFDM). We find that 
these data provide no constraint on the SFDM mass. Therefore, we fix the SFDM mass at $10^{-22}$ eV in this work, 
which also decreases the parameter-space dimensionality and accelerates the numerical evaluation.

Consistent with the convention used in the \textit{Planck} analysis, we include one massive neutrino with $m_{\nu}=0.06$ eV 
and two massless neutrinos to ensure $N_\mathrm{eff}=3.044$ \cite{Froustey_2020,Akita_2020,Bennett_2021}.

\subsection{Datasets}
The following is a list of the various datasets used in our MCMC analysis: 
\begin{itemize}
    \item \textbf{CMB}: The CMB temperature, polarization, and lensing power spectra from the \textit{Planck} 
    2018 data, incorporating both the low-$\ell$ and high-$\ell$ measurements \cite{plk18V,planck2020,plk18VIII}.
    \item \textbf{BAO}: The measurements from the BOSS DR12 $f\sigma_8$ sample include the combined LOWZ and 
    CMASS galaxy samples \cite{Alam_2017, Buen_Abad_2018}, along with low-redshift measurements derived from 6dFGS 
    and SDSS DR7 \cite{19250.x, stv154}.
    \item \textbf{SNIa}: The Pantheon dataset consists of 1048 type Ia supernovae, with redshift values ranging 
    from 0.01 to 2.3 \cite{Scolnic_2018}.
\end{itemize} 
\begin{itemize}
    \item \textbf{SH0ES}: The SH0ES measurement, utilizing the distance ladder method, reports a value of the 
    Hubble constant as $73.04 \pm 1.04$ km/s/Mpc \cite{Riess_2022}.
\end{itemize}
\begin{itemize}
    \item \textbf{DES}: The Dark Energy Survey Year 3 weak lensing and galaxy clustering data provide a 
    Gaussian constraint of $S_8 = 0.776 \pm 0.017$ \cite{PhysRevD.105.023520}.
\end{itemize}

Our baseline dataset consists of a combination of CMB, BAO, and SNIa observations, which we denote as $\mathcal{D}$. 
The SH0ES $H_0$ prior and the DES $S_8$ prior are denoted by $\mathcal{H}$ and $\mathcal{S}$, respectively. 

We adopt the compressed Gaussian prior on $S_8$ from DES Y3 as a proxy for the weak-lensing constraint. While 
the full $3\times2$pt likelihood is the definitive tool for such analyses, its integration with our modified 
Boltzmann solver is technically challenging. Although relevant implementations exist, none can be directly 
combined with our modified dark-sector pipeline without substantial validation -- particularly for the consistent 
integration of the matter power spectrum responses and the joint marginalization over systematic uncertainties. 
Such technical integration, together with the considerable computational cost of full high-dimensional sampling, 
lies beyond the scope of the present proof-of-concept study. We therefore opt for the compressed prior, which 
has been adopted in previous cosmological analyses as a reliable and efficient approximation \cite{PhysRevD.107.083504}, 
as a first-step diagnostic. A full analysis with the complete DES Y3 $3\times2$pt likelihood is reserved to 
future work.    

We first conduct a study using the older data combination mentioned above, and subsequently update 
the cosmological datasets to include the following: 
\begin{itemize}
    \item \textbf{\textit{Planck} NPIPE high-$\ell$} : replaces the 2018 \textsc{plik} likelihood with the 
    \textsc{CamSpec} TTTEEE likelihood from the reprocessed NPIPE pipeline \cite{rosenberg22}, which reduces 
    systematics in TE/EE.
    \item \textbf{\textit{Planck} PR4 lensing}: substitutes the 2018 lensing likelihood with the PR4 release \cite{Carron_2022}, 
    offering higher signal-to-noise and better foreground/beam corrections for robust matter-clustering estimates.
    \item \textbf{ACT DR6 lensing}: a new addition, measuring CMB lensing up to $\ell\sim1000$ with signal-to-noise 
    well above \textit{Planck}, complementary in angular scales \cite{Madhavacheril_2024,Qu_2024}.
    \item \textbf{DESI DR2 BAO}: supersedes the 6dFGS, SDSS-MGS, and SDSS-DR12 combination; covers 
    $0.1<z<4.2$ with multiple tracers, significantly enhancing BAO precision and geometric constraints on 
    $\Omega_m$ and $H_0$ \cite{DESI:2025zpo,DESI:2025zgx}.
    \item \textbf{Pantheon+ supernovae}: replaces the previous Pantheon sample, with $\sim1701$ light curves 
    up to $z=2.26$, improved low-$z$ coverage and systematics, yielding tighter constraints on $H_0$ \cite{Brout:2022vxf}. 
\end{itemize}

This updated combination, especially NPIPE high-$\ell$, PR4 lensing, and ACT DR6 improves both statistical 
precision and systematic control, offering a more rigorous test of our WZDR-SFDM interaction model for 
alleviating the Hubble and $S_8$ tensions. The updated CMB, BAO, and SNIa data are combined with the  
SH0ES and DES data, and the resulting dataset is referred to as ``Updated''.

\subsection{Results}
We present the mean value and 1$\sigma$ parameter constraints for the $\Lambda$CDM model, WZDR model, 
and WZDR+ model, obtained using the baseline dataset $\mathcal{D}$ and the $\mathcal{DHS}$ dataset, 
as shown in Table~\ref{tab:2}. 
\begin{table*}
  \centering
  \caption{The mean values and 1$\sigma$ marginalized parameter constraints for the $\Lambda$CDM model, WZDR 
  model, and WZDR+ model, obtained using various dataset combinations, including $\mathcal{D}$ and 
  $\mathcal{DHS}$, are presented.}
  \label{tab:2}
  \renewcommand{\arraystretch}{1.3}
  \begin{adjustbox}{max width=.9\textwidth}
  \begin{tabular}{|c|c|c|c|c|c|c|}
    \hline
     & \multicolumn{2}{c|}{$\Lambda$CDM}
     & \multicolumn{2}{c|}{WZDR} 
     & \multicolumn{2}{c|}{WZDR+} \\
    \cline{2-7}
     & $\mathcal{D}$ & $\mathcal{DHS}$ 
     & $\mathcal{D}$ & $\mathcal{DHS}$ 
     & $\mathcal{D}$ & $\mathcal{DHS}$ \\
    \hline
    $H_0$&
    $67.68\pm 0.42$ & 
    $68.64\pm 0.35$ & 
    $68.82^{+0.70}_{-1.1}$ & 
    $71.00^{+0.88}_{-0.75}$ & 
    $68.69^{+0.65}_{-1.0}$ & 
    $70.81^{+0.89}_{-0.71}$ \\

    $\Omega_\mathrm{b}h^2$ & 
    $0.02242\pm 0.00014$ & 
    $0.02260\pm 0.00013$ & 
    $0.02251\pm 0.00016$ & 
    $0.02276\pm 0.00015$ & 
    $0.02254\pm 0.00016$ & 
    $0.02279\pm 0.00016$ \\

    $\Omega_\mathrm{dm}h^2$ & 
    $0.11922\pm 0.00093$ & 
    $0.11721\pm 0.00076$ & 
    $0.1223^{+0.0019}_{-0.0029}$ & 
    $0.1248^{+0.0028}_{-0.0024}$ & 
    $0.1219^{+0.0016}_{-0.0027}$ & 
    $0.1240\pm 0.0026$ \\

    $\ln(10^{10}A_\mathrm{s})$ & 
    $3.047\pm 0.014$ & 
    $3.047^{+0.013}_{-0.015}$ & 
    $3.046\pm 0.014$ & 
    $3.043_{-0.017}^{+0.015}$ & 
    $3.048\pm 0.014$ & 
    $3.045\pm 0.015$ \\

    $n_\mathrm{s}$ & 
    $0.9662\pm 0.0036$ & 
    $0.9710\pm 0.0036$ & 
    $0.9708^{+0.0045}_{-0.0054}$ & 
    $0.9795\pm 0.0046$ & 
    $0.9721^{+0.0045}_{-0.0050}$ & 
    $0.9807^{+0.0050}_{-0.0044}$ \\

    $\tau_\mathrm{reio}$ & 
    $0.0561\pm 0.0072$ & 
    $0.0586^{+0.0066}_{-0.0076}$ & 
    $0.0558^{+0.0065}_{-0.0073}$ & 
    $0.0559\pm 0.0073$ & 
    $0.0564\pm 0.0074$ & 
    $0.0565\pm 0.0076$ \\

    $N_\mathrm{IR}$ & 
    ... & 
    ... & 
    $< 0.240$ & 
    $0.43\pm 0.15$ & 
    $< 0.207$ & 
    $0.39\pm 0.14$ \\
    
    $\log_{10}(z_\mathrm{t})$ & 
    ... & 
    ... & 
    $4.30\pm 0.15$ & 
    $4.25^{+0.12}_{-0.19}$ & 
    $4.30\pm 0.16$ & 
    $4.26^{+0.12}_{-0.19}$ \\

    $\log_{10}(\xi)$ & 
    ... & 
    ... & 
    ... & 
    ... & 
    $< -3.14$ (95\% C.L.) & 
    $< -3.36$ (95\% C.L.)\\

    $\sigma_8$ & 
    $0.8095\pm 0.0059$ & 
    $0.8037\pm 0.0055$ & 
    $0.8162^{+0.0071}_{-0.0084}$ & 
    $0.8190\pm 0.0076$ & 
    $0.8159^{+0.0068}_{-0.0080}$ & 
    $0.8180\pm 0.0077$ \\

    $\Omega{}_\mathrm{m}$ & 
    $0.3107\pm 0.0056$ & 
    $0.2982\pm 0.0044$ & 
    $0.3072\pm 0.0058$ & 
    $0.2940\pm 0.0046$ & 
    $0.3075\pm 0.0062$ & 
    $0.2941^{+0.0041}_{-0.0051}$ \\

    $S_8$ & 
    $0.824\pm 0.010$ & 
    $0.8013\pm 0.0082$ & 
    $0.826\pm 0.010$ & 
    $0.8107\pm 0.0093$ & 
    $0.826\pm 0.011$ & 
    $0.8099\pm 0.0098$ \\
    \hline
  \end{tabular}
\end{adjustbox}
\end{table*} 

It is observed that, for the $\mathcal{D}$ dataset, the $H_0$ values for both the WZDR+ and WZDR models are 
larger than those for the $\Lambda$CDM model. When the $\mathcal{H}$ dataset is included, the increase in $H_0$ 
for both models significantly surpasses that for the $\Lambda$CDM model, consistent with previous studies on 
the WZDR model. However, the trade-off for the enhanced $H_0$ in both the WZDR and WZDR+ models is a further 
increase in the $S_8$ parameter, a trend that is observed across all dataset combinations.

Moreover, the $S_8$ parameter constraints for the WZDR+ model are slightly smaller than those for the original 
WZDR model. In the coupled model, the condensation effects of SFDM, along with the momentum exchange between 
dark matter and dark radiation, slightly suppress the growth of cosmic structures, leading to a reduction in 
the $\sigma_8$ parameter.

\begin{figure}
    \centering
	\includegraphics[width=.9\columnwidth]{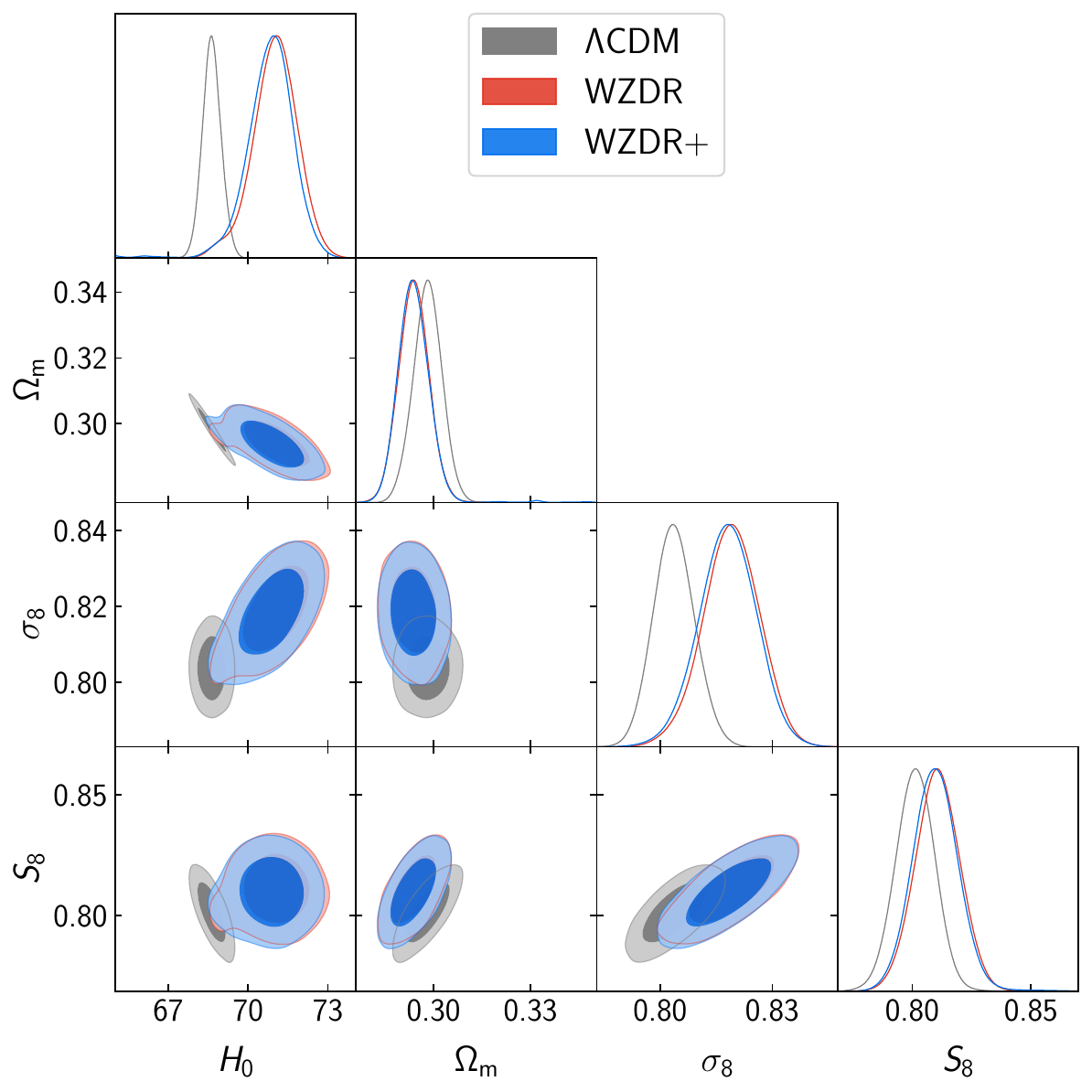}
    \caption{The posterior distribution contours for the parameter constraints of the $\Lambda$CDM model, WZDR 
    model, and WZDR+ model, obtained using the $\mathcal{DHS}$ data combination, are shown. The results for the 
    coupled model closely resemble those of the WZDR model, with only a slight reduction in the value of $S_8$.}
    \label{fig:5}
\end{figure}

However, the distinction between the new model and the WZDR model is minimal. This effect is more clearly 
apparent in the posterior distribution plots of the partial parameters for the three models, as obtained using 
the $\mathcal{DHS}$ data combination, shown in Fig.~\ref{fig:5}.

Table~\ref{tab:3} presents the best-fit $H_0$ and $S_8$ values for the three models under various dataset 
combinations.
\begin{table*}
  \centering
  \caption{The best-fit constraints on the $H_0$ and $S_8$ parameters, along with the $\chi^2_{\min}$ and 
  $Q_{\text{DMAP}}$ values, for the $\Lambda$CDM, WZDR, and WZDR+ models under various dataset combinations.}
  \label{tab:3}
  \renewcommand{\arraystretch}{1.3}
  \begin{adjustbox}{max width=.9\textwidth}
  \begin{tabular}{|c|cccc|cccc|cccc|}
    \hline
     & \multicolumn{4}{c|}{$\Lambda$CDM} & \multicolumn{4}{c|}{WZDR} & \multicolumn{4}{c|}{WZDR+} \\
    \cline{2-13}
     & $H_0$ & $S_8$ & $\chi^2_{\min}$ & $Q_{\text{DMAP}}$ & $H_0$ & $S_8$ & $\chi^2_{\min}$ & $Q_{\text{DMAP}}$ & $H_0$ & $S_8$ & $\chi^2_{\min}$ & $Q_{\text{DMAP}}$ \\
    \hline
    $\mathcal{D}$ 
    & 67.60 & 0.825 & 3819.54 & ...
    & 68.58 & 0.827 & 3820.09 & ...
    & 68.33 & 0.825 & 3820.38 & ...\\

    $\mathcal{DH}$
    & 68.42 & 0.811 & 3841.34 & 4.67$\sigma$ 
    & 71.07 & 0.824 & 3827.00 & 2.63$\sigma$ 
    & 70.69 & 0.820 & 3826.56 & 2.49$\sigma$ \\

    $\mathcal{DS}$
    & 67.76 & 0.811 & 3826.90 & 2.71$\sigma$ 
    & 68.09 & 0.817 & 3825.42 & 2.31$\sigma$ 
    & 68.53 & 0.816 & 3824.39 & 2.00$\sigma$ \\

    $\mathcal{DHS}$
    & 68.63 & 0.802 & 3844.94 & 5.04$\sigma$ 
    & 71.16 & 0.810 & 3833.46 & 3.66$\sigma$ 
    & 70.83 & 0.810 & 3834.65 & 3.78$\sigma$ \\
    \hline
  \end{tabular}
\end{adjustbox}
\end{table*}  
For the datasets $\mathcal{D}$, $\mathcal{DH}$, and $\mathcal{DHS}$, the 
results for $H_0$ and $S_8$ are generally consistent with our previous discussions. However, for the $\mathcal{DS}$ 
dataset, the WZDR+ model outperforms the WZDR model, yielding larger $H_0$ and smaller $S_8$ values.

We present the minimum $\chi^2$ values to compare the goodness of fit for each model. For the dataset combinations 
$\mathcal{D}$, $\mathcal{DH}$, and $\mathcal{DS}$, the WZDR+ model exhibits the smallest $\chi^2_\text{min}$ 
values, indicating the best fit to the data. However, for the $\mathcal{DHS}$ dataset combination, the 
$\chi^2_\text{min}$ value for the WZDR+ model is larger than that of the WZDR model.

To investigate the origin of this discrepancy, we present the $\chi^2$ statistics for the three models using 
the $\mathcal{DHS}$ dataset combination, as shown in Table~\ref{tab:4}. Here, the difference in $\chi^2$ 
values, denoted as $\Delta\chi^2$, represents the contrast between the various models and the $\Lambda$CDM model.
\begin{table}
  \caption{The $\chi^2$ statistical values for fitting the combined $\mathcal{DHS}$ dataset, which 
  including CMB, BAO, SNIa, SH0ES, and DES.}
  \label{tab:4}
  \renewcommand{\arraystretch}{1.2}
  \begin{adjustbox}{max width=.9\linewidth}
  \begin{tabular} { l  c  c c}
     \hline
     Datasets  &  $\Lambda$CDM  & WZDR & WZDR+\\
     \hline
     CMB: & & \\      
     \quad \textit{Planck} 2018 low-$\ell$ TT &22.72 &20.97 &21.04\\        
     \quad \textit{Planck} 2018 low-$\ell$ EE &398.04 &395.81 &396.34\\        
     \quad \textit{Planck} 2018 high-$\ell$\\ 
     \qquad TT+TE+EE &2351.11 &2353.35 &2353.42\\  
     LSS: & & \\      
     \quad \textit{Planck} CMB lensing &9.66 &11.40 &10.52\\             
     \quad BAO (6dF) &0.0257 &0.0796 &0.0828\\             
     \quad BAO (DR7 MGS) &2.24 &2.65 &2.67\\             
     \quad BAO (DR12 BOSS) &6.165 &7.211 &7.275\\         
     SNIa (Pantheon) &1034.73 &1034.77 &1034.79\\    
     SH0ES &17.95 &3.27 &4.54\\    
     DES &2.29 &3.94 &3.98\\    
     \hline 
     $\Delta\chi^2_\mathrm{CMB}$ &... &0.00 &-0.21\\             
     $\Delta\chi^2_\mathrm{BAO}$ &...&1.51 &1.60\\             
     $\Delta\chi^2_\mathrm{SNIa}$ &...&0.04 &0.06\\                
     $\Delta\chi^2_\mathrm{SH0ES}$ &...&-14.68 &-13.41\\                
     $\Delta\chi^2_\mathrm{DES}$ &...&1.65 &1.69\\                
     $\Delta\chi^2_\mathrm{tot}$ &...&-11.48 &-10.38\\      
     \hline         
  \end{tabular}
\end{adjustbox}
\end{table}
We find that although the WZDR+ model provides the best fit to the CMB data, its fit to the BAO and SH0ES data is 
inferior to that of the WZDR model. Overall, the WZDR+ model exhibits a relative disadvantage when compared to the 
WZDR model.

To quantify the deviation between the predicted observables and their direct measurements across the three 
model fits, we calculate the $Q_\mathrm{DMAP}$ values by comparing the best-fit $\chi^2$ valuess, with and without 
the inclusion of direct measurement data $\mathcal{X}$ \cite{PhysRevD.99.043506,SCHONEBERG20221},
\begin{equation}
    Q_\mathrm{DMAP}\equiv\sqrt{\chi^2_\mathcal{DX}-\chi^2_\mathcal{D}},
\end{equation}
the resulting tension metric is expressed in units of Gaussian $\sigma$. 

Table~\ref{tab:3} presents the comparison results between datasets $\mathcal{DH}$, $\mathcal{DS}$, and 
$\mathcal{DHS}$ (which include direct measurements) and the baseline dataset $\mathcal{D}$ for the 
three models. For the $\Lambda$CDM model, the prediction of $S_8$ based on dataset $\mathcal{D}$ exhibits a 
discrepancy of 2.71$\sigma$ relative to dataset $\mathcal{DS}$. The WZDR model reduces this discrepancy to 2.31$\sigma$, 
while the WZDR+ model further decreases it to 2.00$\sigma$. 

A similar trend is observed for the direct measurement of $H_0$, where the comparison between datasets $\mathcal{DH}$ 
and $\mathcal{D}$ reveals that both the WZDR and WZDR+ models produce smaller tension metrics than the $\Lambda$CDM model.

When examining the datasets $\mathcal{DH}$ and $\mathcal{DS}$ individually, the WZDR+ model consistently results in 
smaller $Q_\mathrm{DMAP}$ 
values than the WZDR model. However, when both $\mathcal{H}$ and $\mathcal{S}$ direct measurements are 
included simultaneously (dataset $\mathcal{DHS}$), the WZDR+ model yields a tension metric of 3.78$\sigma$, 
slightly higher than the 3.66$\sigma$ obtained with the WZDR model. This behavior mirrors the trend observed in 
the $\chi^2$ comparisons, indicating that the WZDR+ model performs less favorably than the original WZDR model 
when fitting both direct measurement datasets concurrently.

Finally, we re-analyze all models using the updated CMB, BAO, and SNIa data together with SH0ES and DES, i.e., 
the ``Updated'' dataset defined earlier. 
The the 68\% confidence level marginalized constraints are presented in Table~\ref{tab:5}. 
\begin{table*}
    \centering
    \caption{The 68\% confidence level constraints on the $\Lambda$CDM, WZDR, Uncoupled ($\xi=0$), 
    and WZDR+ models are obtained from the full dataset referred to as the ``Updated'' set in the preceding 
    text. This dataset integrates the latest CMB, BAO, and SNIa observations, as well as measurements from the 
    SH0ES and DES.}
    \label{tab:5}
    \renewcommand{\arraystretch}{1.3}
    \begin{adjustbox}{max width=.9\textwidth}
    \begin{tabular} {|c | c | c | c| c|}
      \hline
        &  $\Lambda$CDM  &  WZDR & Uncoupled  &  WZDR+\\
      \hline
      $H_0$&
      $ 68.53\pm 0.27 $ & 
      $ 71.20^{+0.73}_{-0.61} $ & 
      $ 71.12\pm 0.66 $ & 
      $ 71.08^{+0.73}_{-0.66} $ \\
  
      $\Omega_\mathrm{b}h^2$ & 
      $ 0.02243\pm 0.00012 $ & 
      $ 0.02252\pm 0.00013 $ & 
      $ 0.02254\pm 0.00013 $ & 
      $ 0.02253\pm 0.00013 $ \\
  
      $\Omega_\mathrm{dm}h^2$ & 
      $ 0.11694\pm 0.00060 $ & 
      $ 0.1262^{+0.0025}_{-0.0021} $ & 
      $ 0.1260\pm 0.0023 $ & 
      $ 0.1259^{+0.0026}_{-0.0023} $ \\
  
      $\ln(10^{10}A_\mathrm{s})$ & 
      $ 3.056^{+0.011}_{-0.013} $ & 
      $ 3.048^{+0.011}_{-0.013} $ & 
      $ 3.049\pm 0.012 $ & 
      $ 3.049\pm 0.013 $ \\
  
      $n_\mathrm{s}$ & 
      $ 0.9706\pm 0.0033 $ & 
      $ 0.9779\pm 0.0041 $ & 
      $ 0.9789\pm 0.0041 $ & 
      $ 0.9790\pm 0.0042 $ \\
  
      $\tau_\mathrm{reio}$ & 
      $ 0.0624^{+0.0062}_{-0.0070} $ & 
      $ 0.0571^{+0.0061}_{-0.0074} $ & 
      $ 0.0570\pm 0.0067 $ & 
      $ 0.0571\pm 0.0068 $ \\
  
      $N_\mathrm{IR}$ & 
      $...$ & 
      $ 0.52^{+0.14}_{-0.12} $ & 
      $ 0.51\pm 0.13 $ & 
      $ 0.51^{+0.14}_{-0.13} $ \\
      
      $\log_{10}(z_\mathrm{t})$ & 
      $...$ & 
      $ 4.21^{+0.11}_{-0.14} $ & 
      $ 4.201^{+0.083}_{-0.17} $ & 
      $ 4.201^{+0.088}_{-0.16} $ \\
  
      $\log_{10}(\xi)$ & 
      $...$ & 
      $...$ & 
      $...$ & 
      $ <-3.36 $ (95\% C.L.) \\
  
      $\sigma_8$ & 
      $ 0.8070\pm 0.0048 $ & 
      $ 0.8238\pm 0.0063 $ & 
      $ 0.8234\pm 0.0061 $ & 
      $ 0.8231\pm 0.0064 $ \\
  
      $\Omega{}_\mathrm{m}$ & 
      $ 0.2982\pm 0.0034 $ & 
      $ 0.2948\pm 0.0035 $ & 
      $ 0.2950\pm 0.0033 $ & 
      $ 0.2950\pm 0.0032 $ \\
  
      $S_8$ & 
      $ 0.8045\pm 0.0065 $ & 
      $ 0.8165^{+0.0078}_{-0.0066} $ & 
      $ 0.8164\pm 0.0071 $ & 
      $ 0.8162\pm 0.0073 $ \\
  
      \hline  
      $\Delta\chi^2_\text{min}$ &
      $...$ &
      $-22.16$ &
      $-21.81$ &
      $-18.15$ \\        
      $\Delta\text{AIC}$ &
      $...$ &
      $-18.16$ &
      $-17.81$ &
      $-12.15$ \\        
      \hline      
    \end{tabular}
  \end{adjustbox}
\end{table*}

We find that the updated constraints generally follow the same trends as those obtained with the $\mathcal{DHS}$ 
dataset. For the WZDR and WZDR+ models, $H_0$ increases to $71.20^{+0.73}_{-0.61}$ and $71.08^{+0.73}_{-0.66}$ 
km/s/Mpc, respectively, relative to $\Lambda$CDM, while their $S_8$ mean values both exceed that of $\Lambda$CDM, 
reaching $0.8165$ and $0.8162$. Overall, the two models give very similar results.

To determine whether the reduction in $S_8$ arises from SFDM alone, 
from the momentum coupling, or from both, we also fix $\xi=0$ in an MCMC run and label this model ``Uncoupled''. 
The resulting $H_0$ and $S_8$ lie exactly between those of WZDR and WZDR+, indicating that the interaction 
between SFDM and stepped dark radiation moderately suppresses structure growth and reduces $S_8$.

It should be admitted, however, that the improvement of the WZDR+ model over the original WZDR is only marginal: 
it retains a large $H_0$ while slightly lowering $S_8$, but does not fully resolve the issue. In addition, the 
last two rows of Table~\ref{tab:5} list the minimum $\chi^2$ and the Akaike Information Criterion (AIC) 
\cite{Akaike_1974} values for each model relative to $\Lambda$CDM. Both WZDR and WZDR+ yield smaller $\chi^2$ 
and AIC than $\Lambda$CDM, with the original WZDR model being the best, followed by the ``Uncoupled'' model. 
Thus, from the perspective of $\chi^2$ and AIC, our coupled model performs somewhat worse than the original one.

\begin{figure}
    \centering
	\includegraphics[width=.9\columnwidth]{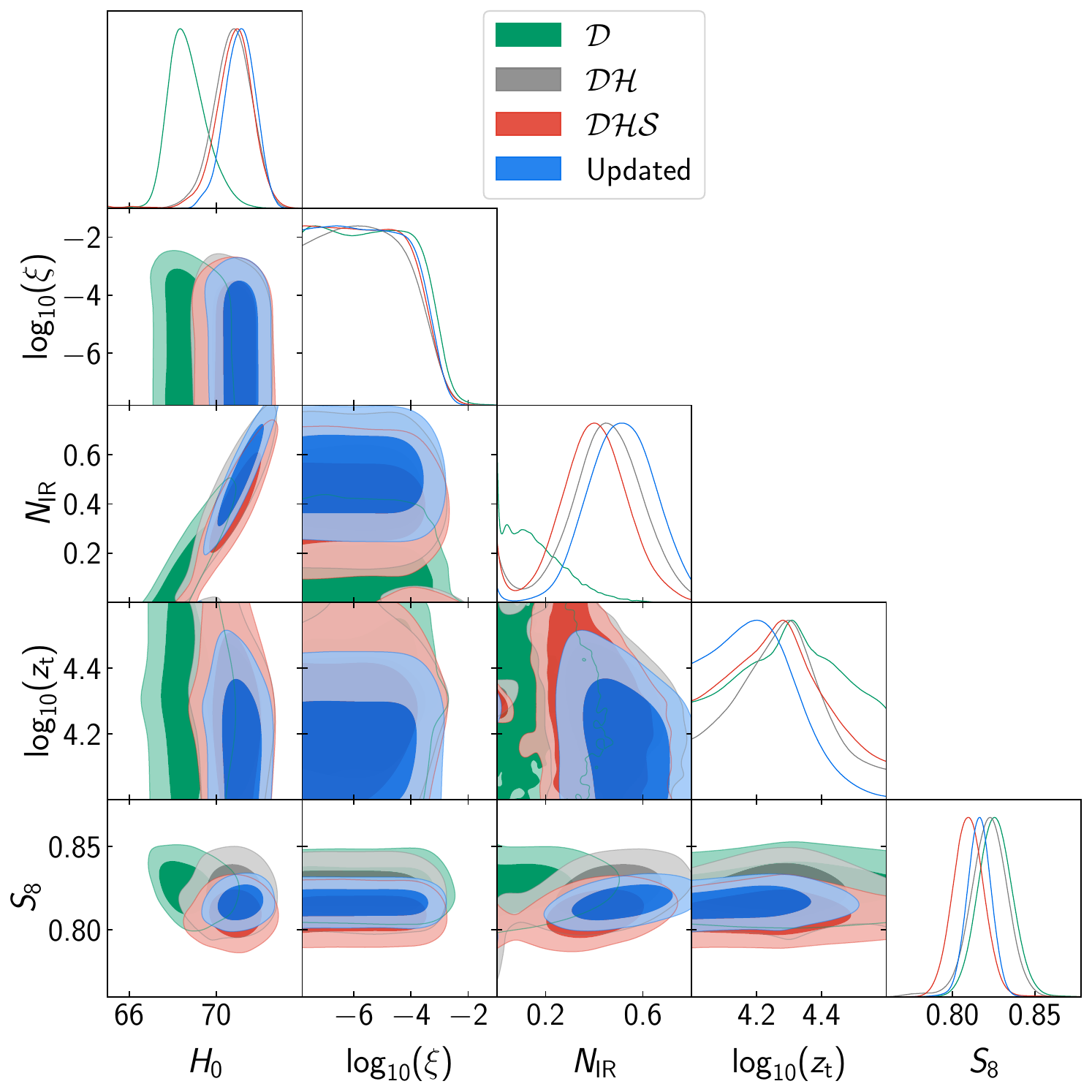}
    \caption{The posterior distributions of selected parameters for the WZDR+ model under various data 
    combinations. The inclusion of datasets $\mathcal{H}$ and $\mathcal{S}$ significantly alters the posterior 
    results for the $H_0$ and $S_8$ parameters. }
    \label{fig:6}
\end{figure}

Figure~\ref{fig:6} presents the posterior distribution contours for selected parameters of the WZDR+ model 
under different data combinations. The complete results for all parameters are provided in Fig.~\ref{fig:7} 
of the Appendix.
We observe that the inclusion of the $\mathcal{H}$ and $\mathcal{S}$ priors significantly improves the 
posterior results for the $H_0$ and $S_8$ parameters when compared to the baseline dataset $\mathcal{D}$. 
Furthermore, the ``Updated'' dataset results in a larger $N_\mathrm{IR}$ parameter, 
which in turn leads to a larger value of $H_0$.

For the coupling parameter $\xi$ in the new model, we find that the constraints from all data combinations 
exhibit a non-Gaussian, long right-tail feature. This is because we sample in $\log_{10}(\xi)$, which restricts 
$\xi>0$ (as motivated in Section~\ref{sec:nr}, where only $\xi>0$ allows the coupling to alleviate $S_8$). For 
robustness, we adopt the 95\% C.L. limits instead of the 68\% ones, and report them in Tables~~\ref{tab:2} and~\ref{tab:5}. 
All datasets yield only upper bounds; in particular, the ``Updated'' dataset gives $\log_{10}(\xi)<-3.36$ (95\% C.L.).    
This indicates that the coupling signal between dark radiation and dark matter is weak, and the differences 
between the WZDR+ model and WZDR model are limited.

\section{Conclusions}
\label{sec:con}
In this paper, we propose a new WZDR+ model by replacing cold dark matter with SFDM 
within the WZDR framework and introducing a pure momentum interaction between SFDM and stepped dark radiation. 
This model is designed to simultaneously address two cosmological tensions: the Hubble tension, which is 
mitigated by the inherent properties of stepped dark radiation, and the $S_8$ tension, which is alleviated 
through the suppression of structure growth driven by SFDM's small-scale condensation and its coupling to dark 
radiation.

We treat the dark radiation as a perfect fluid and introduce the pure momentum coupling between the dark fluid and 
SFDM. Then we investigate the evolution equations for both background and perturbation levels of SFDM and dark 
radiation in the new model, and discuss the impact of interactions on the linear matter power spectrum and the 
CMB temperature power spectrum. 

Subsequently, we perform parameter constraints on the $\Lambda$CDM model, the 
WZDR model, and the WZDR+ model using a variety of cosmological datasets, including CMB, BAO, SNIa, $S_8$ data 
from DES, and $H_0$ data from SH0ES.

We find that the coupled model yields results similar to those of the WZDR model, both of which alleviate the 
Hubble tension compared to the $\Lambda$CDM model. However, this comes at the cost of an increased $S_8$ 
parameter value. Nonetheless, the $S_8$ value of the WZDR+ model is slightly smaller than that of the original 
WZDR model. This is a combined result of the SFDM condensation effect and its interaction with dark radiation in 
the new model.
Therefore, the new coupled model fails to fully resolve the $S_8$ tension, and only marginally alleviates the 
adverse effect of the original WZDR model.

We analyze and compare the minimum $\chi^2$ values and $Q_\text{DMAP}$ values of the three models under 
different data combinations. We find that when the SH0ES and DES data are added individually to the basic 
dataset $\mathcal{D}$ (which includes CMB, BAO, and SNIa data), the WZDR+ model consistently yields the best 
performance, whether evaluated by the minimum $\chi^2$ value or $Q_\text{DMAP}$ value. However, when both SH0ES and DES data are 
included simultaneously, the performance of the WZDR+ model is not as favorable as that of the WZDR model. 
This is because, although the coupled model provides a better fit to the CMB data, its fit to the BAO and 
SH0ES data is not as good as that of the original WZDR model. 

We subsequently update the cosmological datasets to include the most recent CMB data (\textit{Planck} NPIPE 
high-$\ell$ and \textit{Planck} PR4 lensing), ACT DR6 lensing, DESI DR2 BAO, and Pantheon+ supernovae, and 
re-perform MCMC analyses for all models. The new constraints generally follow the same trends as the old ones. 
For $H_0$, the WZDR and WZDR+ models give $71.20^{+0.73}_{-0.61}$ and $71.08^{+0.73}_{-0.66}$ km/s/Mpc, 
respectively, significantly higher than the $\Lambda$CDM value of $68.53\pm 0.27$ km/s/Mpc. However, their 
mean $S_8$ values are $0.8165$ and $0.8162$, both larger than that of $\Lambda$CDM, thus further exacerbating 
the $S_8$ tension.

The WZDR+ model yields a slightly smaller $S_8$ than WZDR. To determine whether this reduction 
originates from the SFDM component itself, from the momentum coupling, or from a combination of both, we 
additionally perform an MCMC analysis with the coupling parameter fixed to $\xi=0$, and denote this case as 
``Uncoupled''. We find that the $H_0$ and $S_8$ results for the ``Uncoupled'' model lie between those of WZDR 
and WZDR+, indicating that both SFDM and its interaction with stepped dark radiation jointly suppress the growth 
of matter structures.

To compare the goodness of fit and the overall performance of the models, we also compute the minimum $\chi^2$ 
and the AIC values. We find that, from both perspectives, the original WZDR model performs the best, followed 
by the ``Uncoupled'' case. Thus, in terms of $\chi^2$ and AIC, the coupled model is disfavored.

For the coupling parameter $\xi$ in the new model, the constraints from all data combinations exhibit 
long-tailed distributions. We therefore adopt the 95\% C.L. limits. All datasets provide only upper bounds on 
$\log_{10}(\xi)$; in particular, the ``Updated'' dataset gives $\log_{10}(\xi) < -3.36$ (95\% C.L.). 
The coupling between dark radiation and dark matter is extremely weak, and the difference between WZDR+ and 
WZDR is marginal.

We have extended the WZDR model by introducing an interaction between SFDM and stepped dark radiation, aiming 
to alleviate both the Hubble tension and the $S_8$ tension. Our numerical analyses show that the new model 
retains the capability of WZDR to raise $H_0$, while slightly reducing $S_8$ and partially mitigating the 
adverse effect of the original model. However, the $S_8$ value in the new model still exceeds that of $\Lambda$CDM; 
thus, the coupled model ultimately fails to resolve the $S_8$ tension. Moreover, from the perspectives of $\chi^2$ 
and AIC, the new model is disfavored compared to the original one.
Although the model does not fully resolve the cosmological tensions, it may serve as a useful baseline 
for future investigations.

\begin{acknowledgments}
    This work was supported by the Fundamental Research Funds for the Provincial Universities of Heilongjiang (No. 2025-KYYWF-ZR0423).
\end{acknowledgments}

\section*{Data Availability}
The data that support the findings of this study are openly available in Figshare at \url{https://doi.org/10.6084/m9.figshare.30819179.v2}.

\appendix*
\section{The full MCMC posteriors}
\begin{figure*}
    \centering
    \includegraphics[width=.9\linewidth]{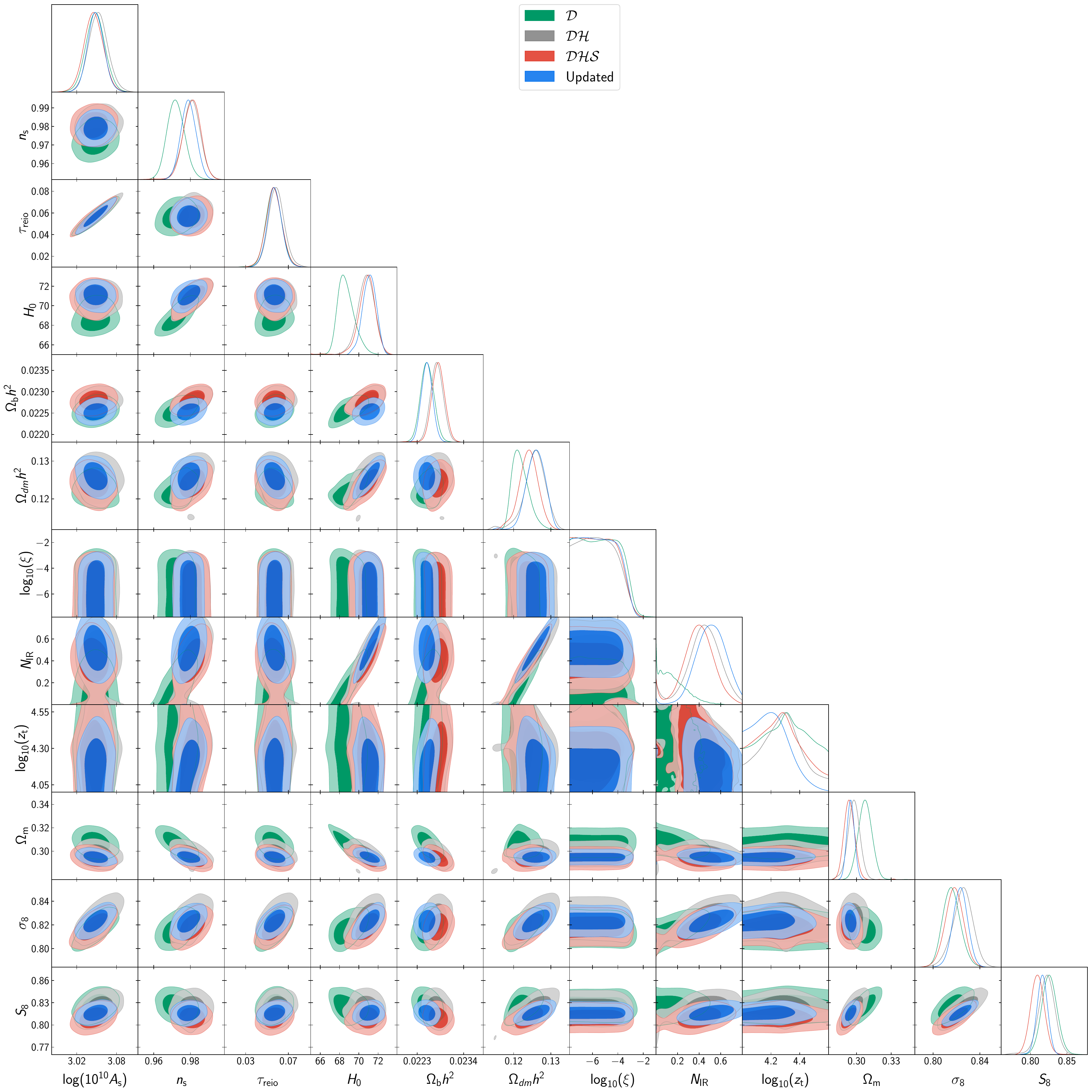}
    \caption{The complete posterior distributions of all parameters for the WZDR+ model under various data 
    combinations, including $\mathcal{D}$, $\mathcal{DH}$, $\mathcal{DS}$, $\mathcal{DHS}$, and ``Updated''.}
    \label{fig:7}
  \end{figure*}

\bibliography{ardrp}

\end{document}